\documentclass[prl,aps,showpacs,preprint]{revtex4}
\usepackage{graphicx}
\usepackage{amsmath,amssymb,revsymb}
\usepackage[colorlinks=true,citecolor=blue]{hyperref}
\usepackage{verbatim}

\begin{document}

\title{Universal van der Waals Physics for Three Ultracold Atoms}

\author{Yujun Wang}
\author{Paul S. Julienne}
\affiliation{Joint Quantum Institute, University of Maryland and NIST, College Park, Maryland, 20742, USA}

\maketitle

{\bf 
Experimental studies with ultracold atoms have enabled major breakthroughs in 
understanding three-body physics, historically a fundamental yet challenging problem.
This is because the interactions among ultracold atoms can be precisely varied using magnetically tunable scattering resonances known as Feshbach resonances~\cite{ChinRev}. 
The collisions of ultracold atoms have been discovered to have many universal aspects near the unitarity 
limit~\cite{Kraemer2006,Berninger2011PRL,Pollack2009,GrossBoth,Wild2012}.
Away from this limit, many 
quantum states are expected to be active during a three-body collision, making the collisional observables practically unpredictable~\cite{LevineMolColli}. 
Here we report a major development in predicting three-body ultracold scattering rates by properly building in the 
pairwise van der Waals interactions plus the multi-spin properties of a tunable Feshbach resonance state characterized by two known dimensionless two-body parameters.
Numerical solution of the Schr{\"o}dinger equation then predicts the three-atom collisional rates without adjustable fitting parameters needed to fit data. 
Our calculations show quantitative agreement in magnitude and feature position and shape across the full range of tuning of measured rate coefficients for three-body recombination~\cite{Kraemer2006} and atom-dimer collisions~\cite{Knoop2009} involving ultracold Cs atoms.
}

The three-body problem has been intriguing ever since the formulation of Newton's laws. In classical mechanics, strongly interacting three-body systems 
often exhibit chaotic behavior, such that the outcome of a three-body collision can be highly sensitive to the details of the interactions. 
The quantum nature of three ultracold atoms makes their collisional properties strongly dependent on the scattering length, a physical entity 
with the units of length that measures a quantum mechanical phase associated with short-range interactions
in a near-zero-energy collision of two atoms.   When a magnetic field $B$ is used to tune this phase by $\pi$ near the pole position $B_0$ of an isolated Feshbach resonance, 
the scattering length $a(B)$ can take on any value in the range $[-\infty,+\infty]$  according to~\cite{ChinRev}
\begin{equation}
a(B)=a_{bg}\left(1-\frac{\Delta}{B-B_0} \right)
\label{Eq_Scatt}
\end{equation} 
where $a_{bg}$ and $\Delta$ are the respective ``background'' scattering length and the width of the resonance.  While universal two- and three-body properties occur when $a(B)$ is much larger than the range of the potential between two atoms~\cite{BraatenRev,WangRev}, it has not heretofore been possible to predict accurate three-body properties when $a(B)$ becomes small.

The sensitivity of three particle dynamics to the details of short-range interactions is dictated by the sign of the effective 
interaction in an abstract coordinate that measures the overall size of a three-body system, the hyperradius $R$~\cite{Delves1959}. 
When the sign is negative, three atoms 
can be accelerated towards small $R$ so that the short-range details become important.  
The sign and strength of the three-body effective interaction is inherently determined by spatial dimension and some three-particle symmetries~\cite{Efimov1970}. 
Remarkably, Thomas's early study indicated that the effective interaction is attractive for three identical bosons, and therefore leads to the so-called Thomas collapse~\cite{Thomas1935} 
when the range of the pairwise forces shrinks to zero. Efimov later discovered an infinity of three-body states 
supported by this effective attraction, whose energies $E_n$ follow a simple geometric scaling: $E_n=E_0 e^{-2n\pi/s_0} (n=0, 1, 2, ...)$, 
where the universal constant $s_0\approx 1.00624$~\cite{Efimov1970}. 
The energy of the ground Efimov state $E_0$ depends on how the effective attractive interaction cuts off at small $R$, and is often 
referred to as the three-body parameter~\cite{BraatenRev,WangRev}. 

In ultracold gases, the formation of Efimov states produces 
features in the three-body recombination loss rate coefficient $L_3$ and atom-dimer relaxation rate $\beta$ that are logarithmically separated in scattering length $a$ ~\cite{BraatenRev,WangRev}; here $L_3$ describes the loss of atomic density $n$ in a sample of ultracold atoms, $dn/dt=-L_3 n^3$.
The positions of such features when $a(B)$ is tuned are connected to the three-body parameter and can therefore manifest three-body short-range physics. 
The remarkable experiments on ultracold Cs atoms have --- for the first time --- evidenced the formation of Efimov states in three-body recombination~\cite{Kraemer2006} and 
atom-dimer relaxation~\cite{Knoop2009}. 
Surprisingly, recent experiments with Cs~\cite{Berninger2011PRL} and other species~\cite{Pollack2009,GrossBoth,Wild2012,Roy2013} have further discovered that the first Efimov resonance occurs universally near 
$a(B) \approx -9 r_{\rm vdW}$, where the van der Waals length $r_{\rm vdW}=(m C_6/\hbar^2)^{1/4}/2$ is purely a two-body quantity defined by the strength of the long-range van der Waals 
interaction $C_6$ and the atomic mass $m$. This length is $~101$ a$_0$ for Cs atoms (a$_0=0.0529$ nm), much longer than the range of normal chemical bonding, but much smaller than the 
size of universal Efimov states. The binding energy of such states is much less than the corresponding van der Waals energy scale $E\lesssim E_{\rm vdW}=\hbar^2/m r_{\rm vdW}^2$, 
which is $E_{\rm vdW}/h\approx 2.7$ MHz for Cs atoms.  

Although arguments have been presented for the existence of a universal three-body parameter independent of the details of short-range interactions~\cite{JWang3BP,Naidon2012b}, 
there is still a question about how a three-body system behaves when it is not in the universal regime where $|a(B)|\gg r_{\rm vdW}$.   
Here we employ a new 2-spin van der Waals model that is both simple and powerful, yet capable of calculating ultracold three-body scattering observables 
across the full range of scattering lengths without knowing the details of short-range interactions.  
Our calculations provide numerical evidence that universality also holds in the small $a(B)$ case previously considered to be non-universal. 
Our model of an isolated Feshbach resonance uses a reduced 2-channel model of the multichannel two body physics~\cite{MiesNygaard} based on two key dimensionless parameters: the background scattering length of the open entrance channel in units of the van der Waals length, $r_{bg}=a_{bg}/r_{vdW}$ and the ``pole strength'' $s_{res}$ of the closed bound state resonance channel~\cite{ChinRev} . 
Both channels are 
represented by a Lennard-Jones 6-12 potential with the same $C_6$~\cite{MiesNygaard}.  The open channel short range part is
selected to yield the known $r_{bg}$ of the resonance and to have $N$ $s$-wave bound states.  
While $N$ can be varied, the three-body properties in our Cs atom case are generally insensitive to its value (see Methods). 
The closed channel differs in magnetic moment from the open channel by $\mu_{dif}$ and has a larger, tunable
separated atom energy. 
The coupling of the open and closed channels is represented by a short-range operator, 
gaussian in our case, chosen so as to reproduce $s_{res} = (4\pi/\Gamma(\frac14)^2) (a_{bg}/r_{vdW}) ( \mu_{dif}\Delta/E_{vdW})$.
Given known values of $r_{bg}$ and $s_{res}$, 
our model contains {\it no} adjustable parameters.

By assuming that each atom has two spin states $|a \rangle$ and $|b \rangle$ with a magnetic moment difference of $\mu_{dif}$, 
we solve the three-body Schr{\"o}dinger equation in hyperspherical coordinates~\cite{WangThesis} with length and energy expressed in reduced units of $r_{vdW}$ and $E_{vdW}$. 
See the Methods Section for details.
This 2-spin model allows us to tune any desired resonance scattering length $a(B)$ and
solve for the three-body bound state energies, $L_3$ coefficient, and atom-dimer relaxation rate coefficient $\beta$ as a function of $a(B)$, given $r_{bg}$ and $s_{res}$.  
Consequently, the three-body scattering in our model is completely determined by the coupled multichannel two-body Feshbach physics together 
with the universal van der Waals physics of the near-threshold ro-vibronic states on the energy scale of $E\lesssim E_{\rm vdW}$~\cite{Gribakin1993,Gao1998,ChinRev}.     

Here we compare our three-body calculations with the well established experimental results for 
Cs atoms~\cite{Kraemer2006,Knoop2009} near the Feshbach resonance centered at $-11.7$G~\cite{ChinRev,Berninger2013}. This resonance has an 
unnaturally large background scattering length with $a_{bg}\approx 1700$a$_0$ $\approx 16.8 r_{\rm vdW}$, and 
$s_{\rm res}\approx 560$ that greatly 
exceeds unity, putting the resonance deep in the regime of broad Feshbach resonances~\cite{ChinRev}. 
Since $L_3$ grows rapidly as $a^4$ for $|a|\gg r_{\rm vdW}$, we follow Ref.~\cite{Kraemer2006} and use the three-body recombination length $\rho_3=[2m/(\sqrt{3}\hbar)L_3]^{1/4}$ to show three-body loss features. 

The effect of three-body short-range physics is typically 
gauged by the following criteria:
(1) the position, width, and $L_3$ magnitude of Efimov resonant features in three-body recombination  for $a<0$;
(2) the position of Efimov intereference minima~\cite{WangRev} in three-body recombination for $a>0$;
(3) the position and magnitude of Efimov resonant features in atom-dimer relaxation for $a>0$.
In the universal limit ($|a|\gg r_{\rm vdW}$)
the above criteria are not independent, since the Efimov features are connected by the universal relations~\cite{BraatenRev,WangRev}. Away from 
this limit, the universal relations are expected to be violated due to contributions from short-range physics.   

\begin{figure}
\includegraphics[scale=0.85]{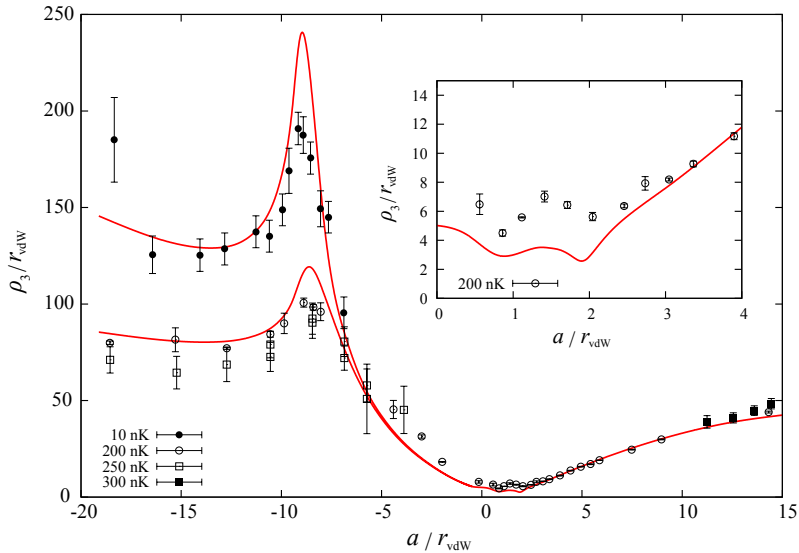}
\caption{
Three-body recombination length $\rho_3$ for Cs atoms near the $-11.7$G Feshbach resonance. The inset magnifies the recombination minima at small $a$ values. 
The solid curves are our thermally 
averaged calculations using the lowest partial wave only at 10 nK (upper curve) and 200 nK (lower curve). The experimental points are shown with error bars~\cite{Kraemer2006}
}
\label{Fig_Recomb}
\end{figure}

Figure~\ref{Fig_Recomb} compares our calculated $\rho_3$ with experiment~\cite{Kraemer2006} at different temperatures. 
The overall agreement is surprisingly good, going 
beyond other theoretical work~\cite{JWang3BP,Schmidt3BP} to explain the position, magnitude and width of the observed Efimov resonance feature at $a\approx -9 r_{\rm vdW}$ without fitting parameters.
The short-range three-body physics is therefore well captured in our model by the first criterion.  
It is also remarkable that such quantitative agreement extends to the features at small $a$. The inset of Fig.~\ref{Fig_Recomb} shows that the first recombination 
minimum shifts away from the universal prediction ($a=0$) to $a\approx 0.79 r_{vdW}$, in good agreement with the experiment. The second minimum is
an Efimov interference minimum at $a\approx 1.9 r_{vdW}$, slightly different from the measured $a\approx 2.1 r_{vdW}$. 
This double minimum structure has not been captured by previous theories~\cite{Previous}.

\begin{figure}
\includegraphics[scale=0.85]{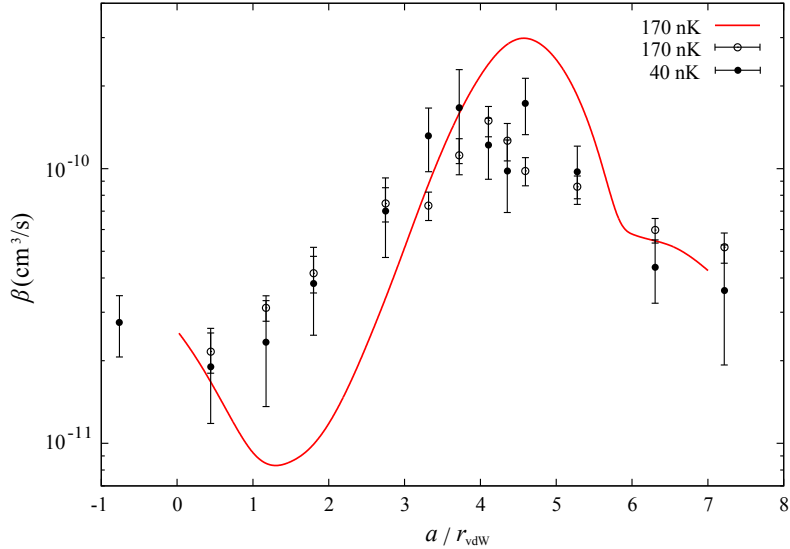}
\caption{ 
Atom-dimer relaxation rate $\beta$ for Cs atoms near the $25$G Efimov resonance.  The solid curve is our thermally 
averaged calculation at 170 nK.
The points with error-bars are the Innsbruck experimental measurements~\cite{Knoop2009} with $a$ mapped to the measured $B$ using updated values for $a(B)$~\cite{Berninger2013}. 
}
\label{Fig_AtomDimer}
\end{figure}

Figure~\ref{Fig_AtomDimer} shows the atom-dimer resonance from Ref.~\cite{Knoop2009} that occurs at $a>0$ where the first excited Efimov trimer bound state crosses the atom-dimer collision threshold.  However, it has been puzzling that the observed position in $a$
is about factor of 2 smaller than predicted by the universal relation. Other theoretical work~\cite{Schmidt3BP} has proposed a strong dependence of
atom-dimer resonances on short-range physics.  But our calculation gives quantitative agreement of magnitude, position, and width in Fig.~\ref{Fig_AtomDimer} between the calculated and measured atom-dimer relaxation rate. Consequently, in this case the two-body parameters $r_{\rm bg}$ and $s_{\rm res}$ and the long-range van der Waals interaction are sufficient to define the ultracold three-body physics without knowing the details of short-range interactions.

\begin{figure}
\includegraphics[scale=0.85]{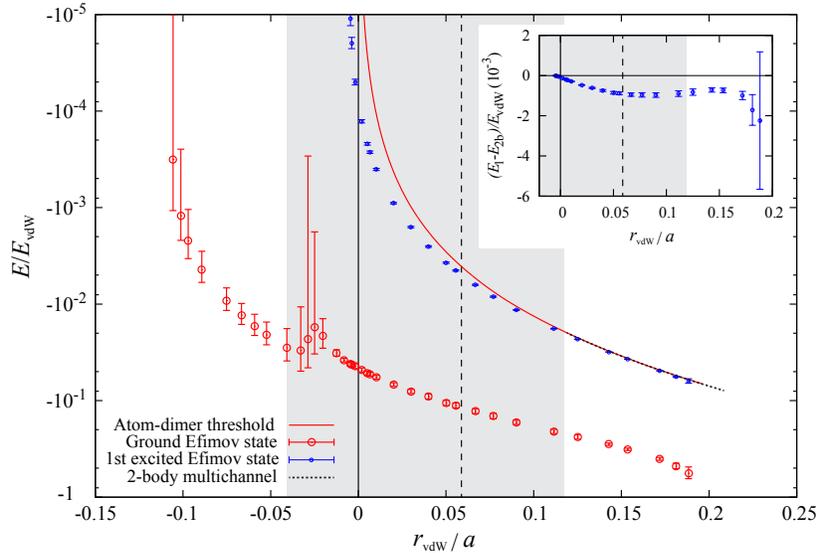}
\caption{
Three-body Efimov energy spectrum for Cs atoms near the $-11.7$G Feshbach resonance. The solid and dotted curves are the atom-dimer thresholds calculated with the 
current two-channel interactions and the realistic, full coupled-channel interactions~\cite{Berninger2013}. The red circles are the energies for the ground 
Efimov state, and the blue circles are for the first excited Efimov state. The vertical bars give the width of the Efimov states due to the decay into the more deeply bound 
atom-dimer channels. The vertical dashed line shows the $r_{\rm vdW}/a$ value where $a=a_{\rm bg}$. The inset shows the energy of the first excited Efimov state 
$E_1$ relative to the atom-dimer threshold $E_{2b}$, where the vertical bars indicate the width of each level.   The grey shaded area shows the range of scattering length that is not accessible to the experiment, which is restricted to be between 0 G and around 48 G, where an additional $d$-wave resonance occurs~\cite{Berninger2013}.
 }
\label{Fig_Spectrum}
\end{figure}

Figure~\ref{Fig_Spectrum} shows
the calculated three-body energy spectrum near the $-11.7$G Feshbach resonance for Cs atoms. The shaded region shows the range of scattering lengths that were not accessible to the experiment.   Nevertheless, to test universality we have calculated 
the energy spectrum using the full range of $a(B)$ for an isolated resonance, and thus can resolve the question~\cite{Schmidt3BP} on how the first Efimov state crosses the atom-dimer threshold.  The connection of the three-body energies 
across the pole of $a$ in Fig.~\ref{Fig_Spectrum} clearly indicates that the ground Efimov state formed around $a=-9 r_{\rm vdW}$ becomes deeply-bound as 
interactions get stronger as $r_{\rm vdW}/a$ changes from negative to positive,  so that this state (red circles) does not cross the atom-dimer threshold (red line).
In our calculation, this ground Efimov state is ``accidentally'' perturbed by a three-body state with $d$-wave character~\cite{JWangDWave} near $r_{\rm vdW}/a \approx -0.04$. 
This $d$-wave state, however, is not universally determined by $a$ (see the Methods section).

The atom-dimer resonance near $a=5 r_{\rm vdW}$ in Fig.~\ref{Fig_AtomDimer} originates from the first excited Efimov state. 
The inset in Fig.~\ref{Fig_Spectrum} shows the binding energy and the width of this first excited Efimov state relative to the atom-dimer threshold. In contrast to
the prediction from the previous universal theory that the three-body binding energy diminishes near the atom-dimer resonance~\cite{BraatenRev,WangRev,Schmidt3BP}, 
our calculation shows that 
the Efimov state broadens and dissolves into the atom-dimer continuum near the resonance at $r_{\rm vdW}/a \approx 0.2$. 
Moreover, the crossing point of its energy with the three-body breakup threshold gives the position of the second Efimov resonance in three-body recombination: 
$a\approx -185 r_{\rm vdW}$, if Eq.~\ref{Eq_Scatt} were to apply here.

To test the universality in our results, we have also calculated the energy of the ground Efimov state by adding a short-range three-body force on the order of the two-body one. The small change in energy, less that 1\% even when $|a|\lesssim r_{\rm vdW}$, indicates that the three-body physics in our study is not affected by short-range three-body forces.
The universality of three-body van der Waals physics is also supported by the continuity of the Efimov state energies across $a=a_{bg}$ in Fig.~\ref{Fig_Spectrum}.
This finding is surprising, considering the drastic change in three-body short-range physics due to the discontinuity in the two-body interactions when $a$ changes from 
$a_{\rm bg}^-$ to $a_{\rm bg}^+$. 
This continuity of the Efimov state energies allows us to test the universal relations~\cite{BraatenRev,WangRev} 
between the three-body recombination peak at $a_-^*\approx -9.3 r_{\rm vdW}$ and the atom-dimer resonance position at $a_+^*\approx 4.7 r_{\rm vdW}$.  The universal relation valid in 
the limit $|a|\rightarrow \infty$ predicts $a_+^*/a_-^*\approx 1.06$. 
whereas the measured ratio is $0.45$~\cite{Knoop2009}.  Our calculated value $a_+^*/a_-^*\approx 0.54$ confirms the measurement in showing that the universal relation is not applicable to the first set of three-body features.

Our three-body model can also be applied to other systems with different $a_{\rm bg}$ and $s_{\rm sres}$. We find that
the universal positions of the three-body Efimov features like those discussed here are more slowly approached 
in the limit of $s_{\rm res}\rightarrow\infty$ with $|a_{\rm bg}|$ larger than that in the case we have presented above. 
For finite, experimentally accessible values of $s_{\rm res}$ there are typically nonnegligible shifts of the three-body features from their universal positions. 
As an example, we have calculated the positions of three-body features near the $550$G broad Cs Feshbach resonance, 
where $a_{\rm bg}\approx 25 r_{\rm vdW}$ and $s_{\rm res}\approx 170$~\cite{ChinRev}. The first recombination minimum and atom-dimer resonance are shifted to $a\approx 2.6 r_{\rm vdW}$ and 
$6.0 r_{\rm vdW}$, respectively, in good agreement with observed values at $2.5 r_{\rm vdW}$~\cite{Ferlaino2011} and $6.0 r_{\rm vdW}$~\cite{InnsbruckUnpub}. While our new 2-spin van der Waals model looks very promising, it will be important to test it on other species and resonances of experimental interest.

\section{Methods}

The total one-body energy $U$ for three atoms $i, j, k$ in a spin state 
$|s_i s_j s_k\rangle$ is 
\begin{equation}
U=-(\mu_{s_i}+\mu_{s_j}+\mu_{s_k})B+u_{s_i}+u_{s_j}+u_{s_k},
\label{Eq_Single}
\end{equation} 
with $|s_{i,j,k}\rangle=|a \rangle$, $|b \rangle$, $\mu_{s_i}$ the magnetic moment, and $u_{s_{i,j,k}}$ the zero-field, single-atom energies.   The interaction $V_{ij}$ between two atoms is represented by a projection into the symmetrized two-body spin basis ${\cal S} |s_i s_j\rangle$: 
\begin{equation}
V_{ij}=\sum_{s_i,s_j,s_k}|s_k\rangle\langle s_k| \otimes |s_i s_j\rangle {\cal S}^\dagger v_{s_i s_j,s_i' s_j'}{\cal S}\langle s_i' s_j'|. 
\label{Eq_Int}
\end{equation}
The diagonal interactions $v_{s_i s_j,s_i s_j}$ are 6-12 Lennard-Jones potentials parameterized by $r_{bg}$ with their depths tuned to have a few 
ro-vibrational bound states. The off-diagonal Gaussian couplings $v_{s_i s_j,s_i' s_j'}$ ($\{s_i s_j\}\neq \{s_i' s_j'\}$) are centered at small distance and are selected consistent with the two-body $s_{res}$ parameter.

The numerical technique we use in this study is similar to those used in 
Refs.~\cite{WangDipoleFermi,WangHetero3BP}, where the hyperspherical coordinates are divided into small and large $R$ regions to apply the Slow Variable Discretization (SVD)~\cite{SVD} 
or the standard adiabatic representation~\cite{WangThesis}. 
Here we use SVD to solve the numerical difficulty from the sharp avoided crossings in the adiabatic representation. 
We symmetrize the three-body wavefunction by imposing its boundary condition in the hyperangular coordinates 
for different spin components, in a similar way to Ref.~\cite{Kokoouline}. 
Due to the existence of deeply-bound atom-dimer thresholds, the Efimov states are in fact resonant states with finite lifetime. The positions and widths of these states are calculated by introducing complex absorbing potentials in the asymptotic atom-dimer channels. 

An important question for the physics of three atoms is the role of the actual two-body ro-vibrational spectrum. Thanks to the universal properties 
of two-body van der Waals interactions~\cite{Gribakin1993,Gao1998}, 
the high-lying two-body ro-vibrational states of open-channel character are well represented in our model. 
Near isolated Feshbach resonances, the property of the closed-channel bound resonant state is almost perfectly reproduced in our model. This is demonstrated 
in Fig.~\ref{Fig_Spectrum} by the agreement between the atom-dimer thresholds from our two-channel model and the full coupled-channel two-body calculations. 
Other ro-vibrational states that have closed-channel character depend strongly on the rotational constant for a particular vibrational state as well as the 
details of the hyperfine couplings. Our three-body calculations indicate that such non-universal states play a minor role in the ultracold physics for three atoms with total energy near the 
three-body breakup threshold, so long as their energies keep away from the threshold for more than a few times of $E_{\rm vdW}$. 

For Cs atoms near the $-11.7$G Feshbach resonance, there are no near-threshold two-body rotational states that couple strongly to the $s$-wave Feshbach state in the range of magnetic field where experimental measurements were taken~\cite{Kraemer2006,Knoop2009}. 
We therefore make the open-channel potential $v_{aa,aa}$ deep enough to hold three $s$-wave bound states so that the 
two-body ro-vibrational states with open-channel character are well represented. The closed-channel 
potential $v_{bb,bb}$ has four $s$-wave bound states and the lowest state is taken as the Feshbach state. This configuration gives 
a physical Frank-Cordon overlap between the Feshbach state and the open-channel scattering state, and also avoids the interruption from unphysical ro-vibrational states.   Two-body closed channel rotational states may accidentally move near the three-body breakup threshold and perturb the Efimov state, as described in Ref.~\cite{JWangDWave}.  An example is the feature with $d$-wave character near $r_{\rm vdW}/a\approx -0.025$.
The positions of such features have no universal connections to $a$.  Except for such accidental features, our calculations have proven to be insensitive to the number of bound states in the potentials.


\begin{thebibliography}{30}
\expandafter\ifx\csname natexlab\endcsname\relax\def\natexlab#1{#1}\fi
\expandafter\ifx\csname bibnamefont\endcsname\relax
  \def\bibnamefont#1{#1}\fi
\expandafter\ifx\csname bibfnamefont\endcsname\relax
  \def\bibfnamefont#1{#1}\fi
\expandafter\ifx\csname citenamefont\endcsname\relax
  \def\citenamefont#1{#1}\fi
\expandafter\ifx\csname url\endcsname\relax
  \def\url#1{\texttt{#1}}\fi
\expandafter\ifx\csname urlprefix\endcsname\relax\def\urlprefix{URL }\fi
\providecommand{\bibinfo}[2]{#2}
\providecommand{\eprint}[2][]{\url{#2}}

\bibitem[{\citenamefont{Chin et~al.}(2010)\citenamefont{Chin, Grimm, Julienne,
  and Tiesinga}}]{ChinRev}
\bibinfo{author}{\bibfnamefont{C.}~\bibnamefont{Chin}},
  \bibinfo{author}{\bibfnamefont{R.}~\bibnamefont{Grimm}},
  \bibinfo{author}{\bibfnamefont{P.}~\bibnamefont{Julienne}}, \bibnamefont{and}
  \bibinfo{author}{\bibfnamefont{E.}~\bibnamefont{Tiesinga}},
  \bibinfo{journal}{Rev.\ Mod.\ Phys.} \textbf{\bibinfo{volume}{82}},
  \bibinfo{pages}{1225} (\bibinfo{year}{2010}).

\bibitem[{\citenamefont{Kraemer et~al.}(2006)\citenamefont{Kraemer, Mark,
  Waldburger, Danzl, Chin, Engeser, Lange, Pilch, Jaakkola, Nagerl
  et~al.}}]{Kraemer2006}
\bibinfo{author}{\bibfnamefont{T.}~\bibnamefont{Kraemer}},
  \bibinfo{author}{\bibfnamefont{M.}~\bibnamefont{Mark}},
  \bibinfo{author}{\bibfnamefont{P.}~\bibnamefont{Waldburger}},
  \bibinfo{author}{\bibfnamefont{J.}~\bibnamefont{Danzl}},
  \bibinfo{author}{\bibfnamefont{C.}~\bibnamefont{Chin}},
  \bibinfo{author}{\bibfnamefont{B.}~\bibnamefont{Engeser}},
  \bibinfo{author}{\bibfnamefont{A.}~\bibnamefont{Lange}},
  \bibinfo{author}{\bibfnamefont{K.}~\bibnamefont{Pilch}},
  \bibinfo{author}{\bibfnamefont{A.}~\bibnamefont{Jaakkola}},
  \bibinfo{author}{\bibfnamefont{H.}~\bibnamefont{Nagerl}},
  \bibnamefont{et~al.}, \bibinfo{journal}{Nature}
  \textbf{\bibinfo{volume}{440}}, \bibinfo{pages}{315} (\bibinfo{year}{2006}).

\bibitem[{\citenamefont{Berninger et~al.}(2011)\citenamefont{Berninger,
  Zenesini, Huang, Harm, N\"agerl, Ferlaino, Grimm, Julienne, and
  Hutson}}]{Berninger2011PRL}
\bibinfo{author}{\bibfnamefont{M.}~\bibnamefont{Berninger}},
  \bibinfo{author}{\bibfnamefont{A.}~\bibnamefont{Zenesini}},
  \bibinfo{author}{\bibfnamefont{B.}~\bibnamefont{Huang}},
  \bibinfo{author}{\bibfnamefont{W.}~\bibnamefont{Harm}},
  \bibinfo{author}{\bibfnamefont{H.-C.} \bibnamefont{N\"agerl}},
  \bibinfo{author}{\bibfnamefont{F.}~\bibnamefont{Ferlaino}},
  \bibinfo{author}{\bibfnamefont{R.}~\bibnamefont{Grimm}},
  \bibinfo{author}{\bibfnamefont{P.~S.} \bibnamefont{Julienne}},
  \bibnamefont{and} \bibinfo{author}{\bibfnamefont{J.~M.}
  \bibnamefont{Hutson}}, \bibinfo{journal}{Phys.\ Rev.\ Lett.}
  \textbf{\bibinfo{volume}{107}}, \bibinfo{pages}{120401}
  (\bibinfo{year}{2011}).

\bibitem[{\citenamefont{Pollack et~al.}(2009)\citenamefont{Pollack, Dries, and
  Hulet}}]{Pollack2009}
\bibinfo{author}{\bibfnamefont{S.~E.} \bibnamefont{Pollack}},
  \bibinfo{author}{\bibfnamefont{D.}~\bibnamefont{Dries}}, \bibnamefont{and}
  \bibinfo{author}{\bibfnamefont{R.~G.} \bibnamefont{Hulet}},
  \bibinfo{journal}{Science} \textbf{\bibinfo{volume}{326}},
  \bibinfo{pages}{1683} (\bibinfo{year}{2009}).

\bibitem[{Gro()}]{GrossBoth}
\bibinfo{note}{N. Gross,Z. Shotan, S. Kokkelmans, and L. Khaykovich, Phys. Rev.
  Lett. 103, 163202 (2009); Phys. Rev. Lett. 105, 103203 (2010).}

\bibitem[{\citenamefont{Wild et~al.}(2012)\citenamefont{Wild, Makotyn, Pino,
  Cornell, and Jin}}]{Wild2012}
\bibinfo{author}{\bibfnamefont{R.~J.} \bibnamefont{Wild}},
  \bibinfo{author}{\bibfnamefont{P.}~\bibnamefont{Makotyn}},
  \bibinfo{author}{\bibfnamefont{J.~M.} \bibnamefont{Pino}},
  \bibinfo{author}{\bibfnamefont{E.~A.} \bibnamefont{Cornell}},
  \bibnamefont{and} \bibinfo{author}{\bibfnamefont{D.~S.} \bibnamefont{Jin}},
  \bibinfo{journal}{Phys.\ Rev.\ Lett.} \textbf{\bibinfo{volume}{108}},
  \bibinfo{pages}{145305} (\bibinfo{year}{2012}).

\bibitem[{\citenamefont{Levine}(2005)}]{LevineMolColli}
\bibinfo{author}{\bibfnamefont{R.~D.} \bibnamefont{Levine}},
  \emph{\bibinfo{title}{Molecular reaction dynamics}}
  (\bibinfo{publisher}{Cambridge University Press}, \bibinfo{year}{2005}).

\bibitem[{\citenamefont{Knoop et~al.}(2009)\citenamefont{Knoop, Ferlaino, Mark,
  Berninger, Schoebel, Naegerl, and Grimm}}]{Knoop2009}
\bibinfo{author}{\bibfnamefont{S.}~\bibnamefont{Knoop}},
  \bibinfo{author}{\bibfnamefont{F.}~\bibnamefont{Ferlaino}},
  \bibinfo{author}{\bibfnamefont{M.}~\bibnamefont{Mark}},
  \bibinfo{author}{\bibfnamefont{M.}~\bibnamefont{Berninger}},
  \bibinfo{author}{\bibfnamefont{H.}~\bibnamefont{Schoebel}},
  \bibinfo{author}{\bibfnamefont{H.~C.} \bibnamefont{Naegerl}},
  \bibnamefont{and} \bibinfo{author}{\bibfnamefont{R.}~\bibnamefont{Grimm}},
  \bibinfo{journal}{Nature\ Phys.} \textbf{\bibinfo{volume}{5}},
  \bibinfo{pages}{227} (\bibinfo{year}{2009}).

\bibitem[{\citenamefont{Braaten and Hammer}(2006)}]{BraatenRev}
\bibinfo{author}{\bibfnamefont{E.}~\bibnamefont{Braaten}} \bibnamefont{and}
  \bibinfo{author}{\bibfnamefont{H.~W.} \bibnamefont{Hammer}},
  \bibinfo{journal}{Phys.\ Rep.} \textbf{\bibinfo{volume}{428}},
  \bibinfo{pages}{259} (\bibinfo{year}{2006}).

\bibitem[{\citenamefont{Wang et~al.}(2013)\citenamefont{Wang, D'Incao, and
  Esry}}]{WangRev}
\bibinfo{author}{\bibfnamefont{Y.}~\bibnamefont{Wang}},
  \bibinfo{author}{\bibfnamefont{J.~P.} \bibnamefont{D'Incao}},
  \bibnamefont{and} \bibinfo{author}{\bibfnamefont{B.~D.} \bibnamefont{Esry}},
  \bibinfo{journal}{Adv.\ At.\ Mol.\ Opt.\ Phys.}
  \textbf{\bibinfo{volume}{62}}, \bibinfo{pages}{1} (\bibinfo{year}{2013}).

\bibitem[{\citenamefont{Delves}(1959)}]{Delves1959}
\bibinfo{author}{\bibfnamefont{L.~M.} \bibnamefont{Delves}},
  \bibinfo{journal}{Nucl. Phys.} \textbf{\bibinfo{volume}{9}},
  \bibinfo{pages}{391} (\bibinfo{year}{1959}).

\bibitem[{\citenamefont{Efimov}(1970)}]{Efimov1970}
\bibinfo{author}{\bibfnamefont{V.}~\bibnamefont{Efimov}},
  \bibinfo{journal}{Phys. Lett. B} \textbf{\bibinfo{volume}{33}},
  \bibinfo{pages}{563} (\bibinfo{year}{1970}).

\bibitem[{\citenamefont{Thomas}(1935)}]{Thomas1935}
\bibinfo{author}{\bibfnamefont{L.~H.} \bibnamefont{Thomas}},
  \bibinfo{journal}{Phys.\ Rev.} \textbf{\bibinfo{volume}{47}},
  \bibinfo{pages}{903} (\bibinfo{year}{1935}).

\bibitem[{\citenamefont{Roy et~al.}(2013)\citenamefont{Roy, Landini,
  Trenkwalder, Semeghini, Spagnolli, Simoni, Fattori, Inguscio, and
  Modugno}}]{Roy2013}
\bibinfo{author}{\bibfnamefont{S.}~\bibnamefont{Roy}},
  \bibinfo{author}{\bibfnamefont{M.}~\bibnamefont{Landini}},
  \bibinfo{author}{\bibfnamefont{A.}~\bibnamefont{Trenkwalder}},
  \bibinfo{author}{\bibfnamefont{G.}~\bibnamefont{Semeghini}},
  \bibinfo{author}{\bibfnamefont{G.}~\bibnamefont{Spagnolli}},
  \bibinfo{author}{\bibfnamefont{A.}~\bibnamefont{Simoni}},
  \bibinfo{author}{\bibfnamefont{M.}~\bibnamefont{Fattori}},
  \bibinfo{author}{\bibfnamefont{M.}~\bibnamefont{Inguscio}}, \bibnamefont{and}
  \bibinfo{author}{\bibfnamefont{G.}~\bibnamefont{Modugno}},
  \bibinfo{journal}{Phys. Rev. Lett.} \textbf{\bibinfo{volume}{111}},
  \bibinfo{pages}{053202} (\bibinfo{year}{2013}).

\bibitem[{\citenamefont{Wang et~al.}(2012{\natexlab{a}})\citenamefont{Wang,
  D'Incao, Esry, and Greene}}]{JWang3BP}
\bibinfo{author}{\bibfnamefont{J.}~\bibnamefont{Wang}},
  \bibinfo{author}{\bibfnamefont{J.~P.} \bibnamefont{D'Incao}},
  \bibinfo{author}{\bibfnamefont{B.~D.} \bibnamefont{Esry}}, \bibnamefont{and}
  \bibinfo{author}{\bibfnamefont{C.~H.} \bibnamefont{Greene}},
  \bibinfo{journal}{Phys.\ Rev.\ Lett.} \textbf{\bibinfo{volume}{108}},
  \bibinfo{pages}{263001} (\bibinfo{year}{2012}{\natexlab{a}}).

\bibitem[{Nai()}]{Naidon2012b}
\bibinfo{note}{P. Naidon, S. Endo, and M. Ueda, arXiv:1208.3912 (2012)}.

\bibitem[{Mie()}]{MiesNygaard}
\bibinfo{note}{F. H. Mies, E. Tiesinga, and P. S. Julienne, Phys. Rev. A 61,
  022721 (2000); N. Nygaard, B. I. Schneider, and P. S. Julienne, Phys. Rev. A
  73, 042705 (2006).}

\bibitem[{\citenamefont{Wang}(2010)}]{WangThesis}
\bibinfo{author}{\bibfnamefont{Y.}~\bibnamefont{Wang}}, Ph.D. thesis,
  \bibinfo{school}{Kansas State University} (\bibinfo{year}{2010}).

\bibitem[{\citenamefont{Gribakin and Flambaum}(1993)}]{Gribakin1993}
\bibinfo{author}{\bibfnamefont{G.~F.} \bibnamefont{Gribakin}} \bibnamefont{and}
  \bibinfo{author}{\bibfnamefont{V.~V.} \bibnamefont{Flambaum}},
  \bibinfo{journal}{Phys.\ Rev.\ A} \textbf{\bibinfo{volume}{48}},
  \bibinfo{pages}{546} (\bibinfo{year}{1993}).

\bibitem[{\citenamefont{Gao}(1998)}]{Gao1998}
\bibinfo{author}{\bibfnamefont{B.}~\bibnamefont{Gao}}, \bibinfo{journal}{Phys.\
  Rev.\ A} \textbf{\bibinfo{volume}{58}}, \bibinfo{pages}{1728}
  (\bibinfo{year}{1998}).

\bibitem[{\citenamefont{Berninger et~al.}(2013)\citenamefont{Berninger,
  Zenesini, Huang, Harm, N{\"a}gerl, Ferlaino, Grimm, Julienne, and
  Hutson}}]{Berninger2013}
\bibinfo{author}{\bibfnamefont{M.}~\bibnamefont{Berninger}},
  \bibinfo{author}{\bibfnamefont{A.}~\bibnamefont{Zenesini}},
  \bibinfo{author}{\bibfnamefont{B.}~\bibnamefont{Huang}},
  \bibinfo{author}{\bibfnamefont{W.}~\bibnamefont{Harm}},
  \bibinfo{author}{\bibfnamefont{H.-C.} \bibnamefont{N{\"a}gerl}},
  \bibinfo{author}{\bibfnamefont{F.}~\bibnamefont{Ferlaino}},
  \bibinfo{author}{\bibfnamefont{R.}~\bibnamefont{Grimm}},
  \bibinfo{author}{\bibfnamefont{P.~S.} \bibnamefont{Julienne}},
  \bibnamefont{and} \bibinfo{author}{\bibfnamefont{J.~M.}
  \bibnamefont{Hutson}}, \bibinfo{journal}{Physical Review A}
  \textbf{\bibinfo{volume}{87}}, \bibinfo{pages}{032517}
  (\bibinfo{year}{2013}).

\bibitem[{\citenamefont{Schmidt et~al.}(2012)\citenamefont{Schmidt, Rath, and
  Zwerger}}]{Schmidt3BP}
\bibinfo{author}{\bibfnamefont{R.}~\bibnamefont{Schmidt}},
  \bibinfo{author}{\bibfnamefont{S.}~\bibnamefont{Rath}}, \bibnamefont{and}
  \bibinfo{author}{\bibfnamefont{W.}~\bibnamefont{Zwerger}},
  \bibinfo{journal}{Euro.\ Phys.\ J. B} \textbf{\bibinfo{volume}{85}},
  \bibinfo{pages}{1} (\bibinfo{year}{2012}).

\bibitem[{Pre()}]{Previous}
\bibinfo{note}{M. D. Lee, Th. K{\"o}hler, and P. S. Julienne, Phys. Rev. A 76,
  012720 (2007); P. Massignan and H. T. C. Stoof, Phys. Rev. A 78, 030701(R)
  (2008); J. P. D'Incao, C. H. Greene, and B. D. Esry, J. Phys. B: At. Mol.
  Opt. Phys. 42 044016 (2009.}

\bibitem[{\citenamefont{Wang et~al.}(2012{\natexlab{b}})\citenamefont{Wang,
  D'Incao, Wang, and Greene}}]{JWangDWave}
\bibinfo{author}{\bibfnamefont{J.}~\bibnamefont{Wang}},
  \bibinfo{author}{\bibfnamefont{J.~P.} \bibnamefont{D'Incao}},
  \bibinfo{author}{\bibfnamefont{Y.}~\bibnamefont{Wang}}, \bibnamefont{and}
  \bibinfo{author}{\bibfnamefont{C.~H.} \bibnamefont{Greene}},
  \bibinfo{journal}{Phys.\ Rev.\ A} \textbf{\bibinfo{volume}{86}},
  \bibinfo{pages}{062511} (\bibinfo{year}{2012}{\natexlab{b}}).

\bibitem[{\citenamefont{Ferlaino et~al.}(2011)\citenamefont{Ferlaino, Zenesini,
  Berninger, Huang, N\"agerl, and Grimm}}]{Ferlaino2011}
\bibinfo{author}{\bibfnamefont{F.}~\bibnamefont{Ferlaino}},
  \bibinfo{author}{\bibfnamefont{A.}~\bibnamefont{Zenesini}},
  \bibinfo{author}{\bibfnamefont{M.}~\bibnamefont{Berninger}},
  \bibinfo{author}{\bibfnamefont{B.}~\bibnamefont{Huang}},
  \bibinfo{author}{\bibfnamefont{H.-C.} \bibnamefont{N\"agerl}},
  \bibnamefont{and} \bibinfo{author}{\bibfnamefont{R.}~\bibnamefont{Grimm}},
  \bibinfo{journal}{Few-Body\ Sys.} \textbf{\bibinfo{volume}{51}},
  \bibinfo{pages}{113} (\bibinfo{year}{2011}).

\bibitem[{Inn()}]{InnsbruckUnpub}
\bibinfo{note}{A. Zenesini, B. Huang, M. Berninger, H. C. N\"agerl, F. Ferlaino, and R. Grimm, unpublished (2013)}.

\bibitem[{\citenamefont{Wang et~al.}(2011)\citenamefont{Wang, D'Incao, and
  Greene}}]{WangDipoleFermi}
\bibinfo{author}{\bibfnamefont{Y.}~\bibnamefont{Wang}},
  \bibinfo{author}{\bibfnamefont{J.~P.} \bibnamefont{D'Incao}},
  \bibnamefont{and} \bibinfo{author}{\bibfnamefont{C.~H.}
  \bibnamefont{Greene}}, \bibinfo{journal}{Phys.\ Rev.\ Lett.}
  \textbf{\bibinfo{volume}{107}}, \bibinfo{pages}{233201}
  (\bibinfo{year}{2011}).

\bibitem[{\citenamefont{Wang et~al.}(2012{\natexlab{c}})\citenamefont{Wang,
  Wang, D'Incao, and Greene}}]{WangHetero3BP}
\bibinfo{author}{\bibfnamefont{Y.}~\bibnamefont{Wang}},
  \bibinfo{author}{\bibfnamefont{J.}~\bibnamefont{Wang}},
  \bibinfo{author}{\bibfnamefont{J.~P.} \bibnamefont{D'Incao}},
  \bibnamefont{and} \bibinfo{author}{\bibfnamefont{C.~H.}
  \bibnamefont{Greene}}, \bibinfo{journal}{Phys.\ Rev.\ Lett.}
  \textbf{\bibinfo{volume}{109}}, \bibinfo{pages}{243201}
  (\bibinfo{year}{2012}{\natexlab{c}}).

\bibitem[{\citenamefont{Tolstikhin et~al.}(1996)\citenamefont{Tolstikhin,
  Watanabe, and Matsuzawa}}]{SVD}
\bibinfo{author}{\bibfnamefont{O.~I.} \bibnamefont{Tolstikhin}},
  \bibinfo{author}{\bibfnamefont{S.}~\bibnamefont{Watanabe}}, \bibnamefont{and}
  \bibinfo{author}{\bibfnamefont{M.}~\bibnamefont{Matsuzawa}},
  \bibinfo{journal}{Journal of Physics B: Atomic, Molecular and Optical
  Physics} \textbf{\bibinfo{volume}{29}}, \bibinfo{pages}{L389}
  (\bibinfo{year}{1996}).

\bibitem[{\citenamefont{Kokoouline and Greene}(2003)}]{Kokoouline}
\bibinfo{author}{\bibfnamefont{V.}~\bibnamefont{Kokoouline}} \bibnamefont{and}
  \bibinfo{author}{\bibfnamefont{C.~H.} \bibnamefont{Greene}},
  \bibinfo{journal}{Phys. Rev. A} \textbf{\bibinfo{volume}{68}},
  \bibinfo{pages}{012703} (\bibinfo{year}{2003}).

\end{thebibliography}

\begin{acknowledgments}
The authors acknowledge the support of an AFOSR-MURI FA9550-09-1-0617, partial support from NSF Grant PHY11-25915, and thank C. H. Greene, J. P. D'Incao and J. Wang for discussions on the method and R. Grimm for providing the original data.
\end{acknowledgments}

YW and PSJ both contributed equally to writing the manuscript.  YW planned the project in consultation with PSJ and implemented the numerical calculations.

\end{document}